\documentclass[aps,pre,superscriptaddress,reprint]{revtex4-1}
\usepackage{bm}
\usepackage{amssymb}
\usepackage{float}	
\usepackage{subcaption}
\usepackage[english]{babel}
\usepackage{tabularx,ragged2e,booktabs,caption}
\usepackage[pdftex]{graphicx}
\usepackage{amsmath}

\usepackage{hyperref}
\usepackage{algpseudocode}
\usepackage{mathtools}
\usepackage{mhchem}

\DeclarePairedDelimiter\abs{\lvert}{\rvert}%
\DeclarePairedDelimiter\norm{\lVert}{\rVert}%

\newcommand{\grad}{\vec{\nabla}}

\begin{document}

\widetext

\title{Theory of Freezing Point Depression in Charged Porous Media}

\author{Tingtao Zhou}
\affiliation{Massachusetts Institute of Technology, Department of Physics}
\author{Mohammad Mirzadeh}
\author{Dimitrios Fraggedakis}
\affiliation{Massachusetts Institute of Technology, Department of Chemical Engineering}
\author{Roland J.-M. Pellenq}
\affiliation{The MIT / CNRS / Aix-Marseille University Joint Laboratory,
``Multi-Scale Materials Science for Energy and Environment" and Massachusetts Institute of Technology, Department of Civil and Environmental Engineering}
\author{Martin Z. Bazant}
\affiliation{Massachusetts Institute of Technology, Department of Chemical Engineering}\affiliation{Massachusetts Institute of Technology, Department of  Mathematics}

\begin{abstract}
 Freezing in charged porous media can induce significant pressure and cause damage to tissues and functional materials. We formulate a thermodynamically consistent theory to model freezing phenomena inside charged heterogeneous porous space. Two regimes are distinguished: free ions in open pore space lead to negligible effects of freezing point depression and pressure. On the other hand, if nano-fluidic salt trapping happens, subsequent ice formation is suppressed due to the high concentration of ions in the electrolyte. In this case, our theory predicts that freezing starts at a significantly lower temperature compared to pure water. In 1D, as the temperature goes even lower, ice continuously grows, until the salt concentration reaches saturation, all ions precipitate to form salt crystals, and freezing completes. Enormous pressure can be generated if initial salt concentration is high before salt entrapment. We show modifications to the classical nucleation theory, due to the trapped salt ions. Interestingly, although the freezing process is enormously changed by trapped salts, our analysis shows that the Gibbs-Thompson equation on confined melting point shift is not affected by the presence of the electrolyte.
\end{abstract}

\date{\today}

\maketitle

\section{Introduction}
Freezing tolerance is necessary for materials that experience cold conditions, however its mechanisms are not yet clear in various contexts. For example, freeze-thaw damage is one of the biggest threats to cement and concrete in cold areas. Although de-icing salt lowers the melting point of snow and ice on the roads, it actually makes the damage worse~\citep{ farnam2014acoustic, farnam2014measuring}. The conventional thinking of water expansion upon freezing causing damage~\citep{PCAwebsitefreezethaw} contradicts this fact, and the real mechanism of freezing damage is subject to more careful investigations. 
Frost heave damage in soils has been discussed~\cite{wettlaufer2006premelting}, and theorists achieved successes in explaining the deformation of saturated soils due to the dynamics of pre-melted liquid and its coupling with the solid. However, its applicability to cement is questionable, as the cohesion of cement paste nano-particles are much stronger than the capillary forces~\cite{zhou2019capillary,zhou2019multiscale}. 
Biological materials also exhibit remarkable freezing endurance: Human embryos can be safely preserved using liquid nitrogen at -200~$^{\circ}$C~\citep{abdelhafez2010slow, saragusty2011current, wakchaure2015review, rienzi2017oocyte}. Bacteria, some arctic insects and other primitive forms of life can survive extremely cold weathers of -60$\sim$-100~$^{\circ}$C~\citep{miller1987extreme, ring1981physiology}. Animals on the higher branches of the ``evolution tree'' such as amphibians and reptiles show moderate freezing tolerance around -10~$^{\circ}$C during winter hibernation~\citep{layne1987freeze, storey1988freeze, storey2001hibernation, storey2006reptile}. Perennial plants also survive freezing weathers in winter~\citep{andrews1996plants, guy1990cold}. Their amazing capabilities of freezing tolerance are usually associated with anti-freezing proteins~\citep{storey2013molecular, janmohammadi2015low, hoshino2003antifreeze, wen2014protective}. Here our theory proposes a more general potential physical mechanism contributing to freezing tolerance and damage.
The thermodynamics of freezing is complicated by the existence of salt ions and charged pore surface. Nucleation mechanisms have been studied both theoretically and numerically\citep{matsumoto2002molecular}. The classical nucleation theory (CNT) predicts the critical size of nuclei and the nucleation rate at the beginning of freezing~\citep{debenedetti1996metastable}. Nucleation inside a charged pore filled with electrolyte requires modifications to the classical theory. 
While the influence of salt on the bulk solution freezing has been extensively studied~\citep{ debye1923theorie}, the role of salt ions in the confined freezing of water has yet to be explored. Supercooling and freezing point depression due to salt  are critical processes for freezing tolerance in heterogeneous porous media.
In this paper, we specifically distinguish the regimes of free ions and trapped
ions, and propose nano-fluidic salt trapping mechanisms for heterogeneous
porous media. We present a thermodynamically consistent theory to predict
freezing point depression, pressures and modifications to CNT equations, and discuss the Gibbs-Thomson effect of melting point.  

\section{Theory}

\begin{figure}
 \includegraphics[width=1\linewidth]{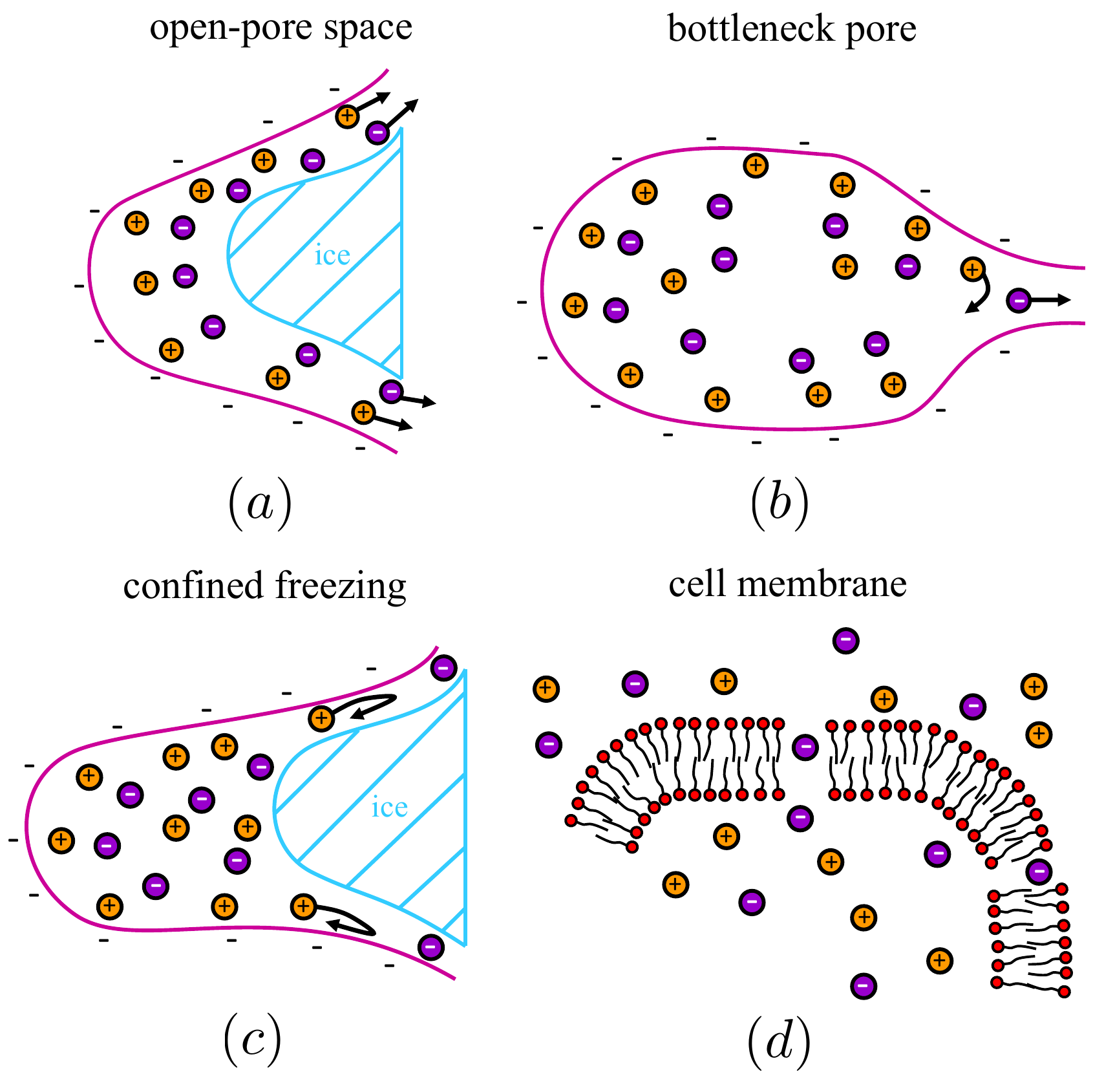}
 \caption{
  \label{fig:figure1}
  Free ions v.s. nano-fluidic salt trapping.
  (a) when the freezing pore is well connected to the reservior, the ions move through the pathway and not constrained inside the pore. This case is referred to as ``free'' ions. (b) when the freezing pore is only connected via bottlenecks, geometric constraints make the solvated ions difficult to transport through the bottleneck. (c) when heterogeneous freezing happens and the freezing pore is blocked by ice formed in its larger neighbors, only a thin (< nm size) layer of lubricating liquid layer exists as channels to connect to the reservior. Transport of the ions deviates from bulk transport behavior and they will be hindered due to the highly charged surface of the narrow liquid channels. (d) the ion and water channels on a membrane can actively control these transport processes, hence controlling the salt concentration inside.
   }
\end{figure}

Let's consider a charged pore space filled with electrolyte. When temperature lowers, the liquid state becomes thermodynamically unfavorable and is inclined to solidify. If the pore space is well connected to a large reservoir that can accommodate the salt ions and excess water molecules, then once freezing begins all the water molecules in this pore should turn into solid ice, except for an interfacial layer of liquid ($<$1~nm thickness) remaining between the pore surface and ice core~\citep{denoyel2002simple}, and the salt ions will escape into the large reservoir. Below in Section.\ref{sec:OCP} we refer to this situation as free salt ions, where only a certain number of counter-ions remains in the pore to balance the pore surface charge, and to contribute to freezing point depression and pressure (see Fig.1 (a)).
On the other hand, if the freezing pore is disconnected from the external reservoir, such that the salt ions experience strong resistance to removal from the pore, then the crowding salt ions can inhibit ice growth (see Fig.1 (b-d)). In this case, ice formation generates significant pressure that is transmitted to the solid matrix of the porous medium. Eventually, the resistance to freezing from salt ions will abruptly end once the electrolyte volume is squeezed by ice so that it reaches concentration saturation. All salt ions will precipitate into crystals at this point and the entire pore space will freeze, except for the remaining lubricating liquid layer mentioned above.

The latter case of effectively disconnected pore may be caused by several underlying mechanisms such as: 1) the pore is only connected through a narrow bottleneck to larger space (Fig.1(b)); 2) the pore is initially connected through another big pore, however when temperature lowers the big pore is frozen first and only a narrow channel of lubricating liquid remains (Fig.1(c)); 3) active control of water/ion transport by channels on a cell membrane (Fig.1(d)). In 1) and 2), when surface charge on the narrow channels are high, co-ions will tend to be excluded from the channel and hence cannot exit from the freezing pore; counter-ions remain to maintain overall charge neutrality. If the solvation size of the ions is comparable or larger than the narrow channels, the ions are geometrically trapped. In all these and other similar situations, the solvated salt ions exit the pore over a longer transport timescale to get through the highly charged and narrow channels of nm size, so that the freezing process occuring on a shorter timescale will be significantly affected by their existence in the pore. We distinguish water molecules as ``solvated'' for those in the ion solvation shell and ``free'' for those not. As with the free water molecules, the solvated water molecules are electrically neutral and of very small size ($\sim$ 3~$\mbox{\AA}$), hence we safely assume that all ``free'' water molecules exchange between the freezing pore and the external reservoir, even when salt is trapped.
Below in Section.\ref{sec:trapped ions} we refer to this situation as nano-fluidic salt trapping. This idea of nano-fluidic salt trapping has led to the development of several nano-fluidic devices, such as electro-osmotic micropumps\citep{zeng2001fabrication}, nano-fluidic diodes and bipolar transistors \citep{daiguji2005nanofluidic, yossifon2006electro, yossifon2009rectification}, and nanofluidic ion separators \citep{gillespie2013separation}.

Throughout this work, we adopt a continuum description for both the electrolyte and the ice domains. The general form of free energy functional is
\begin{equation}
 \begin{split}
  F_{tot} & = F_{liquid} + F_{solid} + F_{interface} \\
  & = \int_{V_s} dV \left( \mu_s - \mu_l - \frac{\epsilon_{s}}{2} \norm{\grad \phi}^2 \right)  \\
  & + \int_{V_l} dV \left[
  g(\{c_i\}) + \rho\phi - \frac{\epsilon_l}{2} \norm{\grad \phi}^2  \right] \\
  & +  \sum_{j=s, l, sl} \int_{S_{j}} dS \left( \gamma_j+ q_j\phi \right) \\
 \end{split}
 \label{eq:Fgen}
\end{equation}
where the integrations are performed over the volumes of the solid ($V_s$) and the liquid ($V_l$) with permittivities $\epsilon_s$ and $\epsilon_l$, respectively, and over surfaces of the solid-liquid interface ($S_{sl}$), the liquid-pore interface ($S_l$) and the solid-pore interface ($S_s$), with corresponding surface charge densities, $q_{sl}$, $q_{l}$ and $q_{s}$ and interfacial tensions, $\gamma_{sl}$, $\gamma_l$ and $\gamma_s$; $\mu_s - \mu_l$ is the bulk chemical potential difference between solid and liquid phases; $-\grad \phi$ is the electric field; $g(\{c_i\})$ the non-electric part of homogeneous liquid electrolyte free energy;  $c_i$ the concentration of ion species $i$ having charge $z_i e$; and $\rho = \sum_i z_i e c_i$ the net charge density, assumed to be negligible in the solid phase.  We focus on situations of complete wetting by the liquid,  $\gamma_s-\gamma_l \gg \gamma_{sl}$, in which case we can neglect $S_s$ and assume $S_l$ covers the entire pore surface.

We solve the model for isotropic symmetric pore spaces in d-dimension (d=1,2,3), and assume a smooth ice/liquid interface respecting the same symmetry of the pore geometry. One significant convenience from the symmetries is that no electric field should penetrate into the ice core in the middle of the pore, so that the electric field energy term inside $V_s$ can be neglected. To show that, one can draw a Gauss surface surrounding the ice, or placed concentrically inside the ice core, with the same geometric symmetry. The total flux of the E-field through this surface should be 0 since we assume there is no net charge in the ice
\begin{equation}
 \int_{S_{\text{Gauss}}} \vec{E} \cdot d\vec{S} = 0
\end{equation}
By symmetry each infinitesimal element of $\vec{E} \cdot d\vec{S}$ should possess the same value regardless of the orientation of the normal direction $d\vec{S}$, leading to $\vec{E}\equiv 0$.

To determine the equilibrium state of the system under a given pore radius, $R$, pore surface charge density, $eq$, and initial salt concentration, $c_0$, we minimize the total free energy functional Eqn.\ref{eq:Fgen}, w.r.t. the position of ice/liquid interface, denoted as $r$.
For each fixed $r$ value, the electric field and ion density distribution inside the liquid phase should also minimize the free energy.
For simplicity, we choose the entropy of an ideal gas
\begin{equation}
 g(c(\phi)) = k_B T c (\ln(cv)-1)
\end{equation}
so that the equilibrium ion concentration follows a Boltzmann distribution (for other entropy such as a lattice gas model corresponds to a Fermi-like distribution).
The variational principle, ${\delta F}/{\delta \phi}=0$, leads to the Poisson-Boltzmann (PB) equations~\citep{gouy1910constitution, chapman1913li, andelman1995_PBforces}. We assume that ice always forms concentrically at the center of the pore from homogeneous nucleation. Additionaly, we assume the ice surface to be neutral and use $q$ to denote the number density of surface charges on the pore wall.


\subsection{Free Ions Limit: One-Component Plasma}
\label{sec:OCP}
For one-component plasma (OCP) the PB equation reads
\begin{equation}\label{eq:OCP_general}
 \nabla^2 \Phi + \kappa_{ocp}^2 e^{-\Phi} = 0
\end{equation}
where $\Phi=Z e \beta \phi$ is the dimensionless potential, $\beta=1/k_B T$ and $\kappa_{ocp}^2 = Z \beta e^2/\epsilon l_B$, where $l_B = \beta Z^2 e^2/4\pi\epsilon_l$ is the Bjerrum length. Additionaly, the boundary conditions for $\Phi$ are
\begin{equation}
 \begin{split}
  \Phi'(r/l_G) & = 0 \\
  \Phi'(R/l_G) & = -\frac{Z e^2\beta l_G}{\epsilon} q
 \end{split}
\end{equation}
We scale all the spatial coordinates with the Gouy-Chapman length $l_{G}= \frac{4\pi\epsilon_l k_B T}{Z e^2 q_l}$ as $\tilde{x}=x/l_G$. For OCP the value of the reference electric potential is only determined by the prefactor of $e^{-\Phi}$. Here, we choose the nondimensional prefactor $\kappa_{ocp}$ in terms of these characteristic length scales.

\subsubsection{d=1: OCP in a slit pore geometry}
In the one-dimensional case Eq.~\ref{eq:OCP_general} becomes
\begin{equation}
 \Phi'' + {\kappa_{ocp}^2} e^{-\Phi} = 0
\end{equation}
and the surface of the ice is located at $r=0$.
This equation is integrable noticing that
\begin{equation}
 \begin{split}
  \Phi'' \Phi' + \kappa_{ocp}^2 e^{-\Phi} \Phi' & = 0 \\
  \frac{d}{dz} \left[ \frac{1}{2} (\Phi')^2 - \kappa_{ocp}^2 e^{-\Phi}
  \right] & = 0 \\
  \frac{1}{2} (\Phi')^2 - \kappa_{ocp}^2 e^{-\Phi} & = -\tilde{P}
 \end{split}
\end{equation}
For $\tilde{P}>0$ the solution reads
\begin{equation}
 \phi = \frac{1}{Ze\beta}\ln \frac{\kappa_{ocp}^2}{\vert\tilde{P}\vert}\sin^2\left( \sqrt{\frac{\vert\tilde{P}\vert}{2}} \tilde{x} + \theta_0 \right)
\end{equation}
where the constants $\tilde{P}$ and $\theta_0$ are determined by the boundary conditions
\begin{equation}
 \begin{split}
  \phi'(\tilde{x}=0) & = \frac{\sqrt{2\vert\tilde{P}\vert}}{Ze\beta} \cot (\theta_0) = 0 \\
  \phi'(\tilde{x}=R/l_G) & = \frac{\sqrt{2\vert\tilde{P}\vert}}{Ze\beta} \cot \left(\sqrt{\frac{\vert\tilde{P}\vert}{2}} R + \theta_0\right) = - e q l_G/\epsilon
 \end{split}
\end{equation}
so the final expression of electric potential reads
\begin{equation}
 \begin{split}
  & \phi  = \phi_0 - \frac{1}{Ze\beta}\ln \vert \tilde{P} \vert \cos^2\left( \sqrt{\frac{\vert \tilde{P} \vert}{2}} \tilde{x} \right) \\
  & \sqrt{\frac{\vert \tilde{P} \vert}{2}} \frac{R}{l_G} \tan \left( \sqrt{\frac{\vert \tilde{P} \vert}{2}} \frac{R}{l_G} \right) = \frac{Zq\beta e^2}{2\epsilon} = \frac{2\pi R}{l_G} \\
 \end{split}
 \label{eqn:sol-planar-PB-1component}
\end{equation}
Substituting the above solution into the total free energy (per unit area $A$) of the system
\begin{equation}
 \frac{F}{A} = [\mu_s - \mu_l] x + \int_r^R dx \left(
 g(c(\phi)) + \frac{\epsilon}{2} (\phi')^2
 \right)
\end{equation}
where minimization of this form yields
\begin{equation}\label{eq:min_FE_OCP}
 -\frac{\partial}{\partial x}\frac{F}{A} =
 -[\mu_s-\mu_l] - \frac{k_B T}{4\pi l_B l_G^2} \tilde{P} = 0
\end{equation}
The Gibbs-Helmholtz equation relates the bulk freezing enthalpy to: i) the latent heat of bulk water $Q$, ii) the difference of heat capacity between water and ice $\Delta c_p$, iii) the bulk freezing point $T_0$, iv) and the freezing point depression $\Delta T=T-T_0$ as
\begin{equation}
 \mu_s - \mu_l = \left( Q\frac{\Delta T}{T_0} - \Delta c_p \frac{\Delta T^2}{T_0} \right)
\end{equation}
Combining this result with Eq.~\ref{eq:min_FE_OCP} we arrive at a relation between $\tilde{P}$, $Q$, $\Delta T$, $T_0$ and $\Delta c_p$ as follows
\begin{equation}
 \tilde{P} = \frac{4\pi l_B l_G^2}{k_B T} \left\vert Q\frac{\Delta T}{T_0} - \Delta c_p \frac{\Delta T^2}{T_0} \right\vert
 \label{eqn:ocp-pressure}
\end{equation}
The physical meaning of $\tilde{P}$ is shown to be the dimensionless pressure. When no curvature effects are present, freezing point depression is achieved via the presence of counter-ions. More specifically,
if $q$ increases, in order for the ice not to melt, i.e. the ice front should remain at $r$, $\tilde{P}$ has to increase (Eqn.~\ref{eqn:sol-planar-PB-1component}) resulting in a decrease of the freezing temperature, i.e. $\Delta T$ becomes even more negative.
A quick order of magnitude estimate shows that
when $2\pi d\gg l_G$ (for free water $\epsilon_r=80$, $q=1$~nm$^{-2}$, $l_B\approx 0.6$~nm, $l_G\approx 1.66$~nm, with $R\sim 6$~nm we have $2\pi d/l_G\approx 22.7$). Solving Eqn.~\ref{eqn:sol-planar-PB-1component} we arrive at 
\begin{equation}
 \begin{split}
  P_{ele} & \approx \frac{\pi k_B T}{8 R^2 l_B} \sim
  P_{freeze} \approx -Q \frac{\Delta T}{T_0} \\
  \Delta T & \sim -0.1~K \left(\frac{T_0}{273~\mathrm{K}}\right) \left(\frac{10~\mathrm{nm}^3}{R^2 l_B}\right)
 \end{split}
\end{equation}
Fig.\ref{fig:OCP-free-energy-profiles} (a)(d) depict typical free energy functions at T=270~K and 230~K, respectively, for a pore of radius $R$=5~nm and surface charge density $q=$1~nm$^{-2}$. Since d=1 there are no contributions from the surface energy. The competition between OCP and ice freezing enthalpies determines whether the pore is frozen or not. As analyzed previously, as soon as the temperature drops about 1~K below the bulk freezing point, the entire pore freezes. Pressure values can be estimated by using Eqn.\ref{eqn:ocp-pressure}.

\subsubsection{d=2,3: OCP in a cylindrical/spherical pore symmetry}
Under cylindrical/spherical symmetry, $\partial_{\theta/\phi} \Phi=0$, respectively, with $\theta$ and $\phi$ denoting the azimuthal angles. After separation of variables, the radial part of the non-linear dimensionless Poisson-Boltzmann equation reads
\begin{equation}
 \Phi'' + \frac{d-1}{\tilde{x}} \Phi' + e^{-\Phi} = 0
 \label{eqn:OCP-radial-ode}
\end{equation}
with
\begin{equation}
 \begin{split}
  \Phi & = \beta Z e \phi + \ln\left( \frac{\epsilon k_B T}{Z^2 e^2 c_0 l_G^2}\right) \\
  & = \beta Z e \phi + \ln\left( \frac{q^2 e^2}{(4\pi)^2 \epsilon_r\epsilon_0 k_B T c_0} \right) \\
  & = \beta Z e \phi - \ln\left( 4\pi c_0 l_B l_G^2 \right)
 \end{split}
\end{equation}
where surface charge density is $-e q<0$ (assumed negatively charged pores). The counter-ion concentration is
\begin{equation}
 c(x) = c_0 e^{-\beta Z e \phi} = e^{-\Phi} \frac{1}{4\pi l_B l_G^2}
\end{equation}
where $c_0$ is defined as the concentration when $\phi=0$.

In the remaining of the text {\bf we drop the hat symbol from all dimensionless quantities}, so that $\tilde{x}=x/l_G \rightarrow x$, $\tilde{r}=r/l_G \rightarrow r$, $\tilde{R}=R/l_G \rightarrow R$.
The boundary conditions for the higher-dimensional case read
\begin{equation}
 \begin{split}
  \Phi'(x=r) & = 0 \\
  \Phi'(x=R) & = \frac{Z e}{k_B T} \partial_x \phi (R) = -4\pi l_G
 \end{split}
 \label{eqn:ocp-BC-cyl-sph}
\end{equation}

For OCP, charge neutrality can be checked by integrating the charged surface boundary condition
\begin{equation}
 \begin{split}
  & l_G^{d}\int_{r}^{R} S(d) x^{d-1}dx \frac{Z}{l_B l_G^2} e^{-\Phi(x)} = S(d) R^{d-1} l_G^{d-1} q \\
  & \int_{r}^{R} x^{d-1}dx e^{-\Phi(x)} =  S(d) R^{d-1}
 \end{split}
\end{equation}
where $S(d)$ is the dimensionless coefficient of surface area of a hyper-sphere ($A_R(d)=S(d)R^{d-1}$) in d dimension: $S(1)=1, S(2)=2\pi, S(3)=4\pi$.

A general change of variable $z(x)$ yields
\begin{displaymath}
 \begin{split}
  \Phi' & = \partial_z \Phi \partial_{x} z = z' \partial_z \Phi \\
  \Phi{''} & = z^{''} \partial_z\Phi + {z'}^2 \partial_z^2 \Phi
 \end{split}
\end{displaymath}
The prime $'$ for derivative w.r.t $x$. Now Eqn.\ref{eqn:OCP-radial-ode} is
\begin{equation}
 \left( {z'}^2 \partial_z^2 \Phi + e^{-\Phi} \right) + \left[ z'' + (d-1)z'/x \right] \partial_z \Phi = 0
\end{equation}
One way to simplify the differential equation is to eliminate the $\partial_z\Phi$ term. This is done by setting its prefactor equal to zero as
\begin{equation}
    x z'' + (d-1)z' = 0 \rightarrow \left( x^{(d-1)}z' \right)' = 0
    \label{eq:mapping_deq}
\end{equation}
Eq.~\ref{eq:mapping_deq} consists of a differential equation which evaluates the mapping between the physical coordinates $r$ and the reference space $z$. In the case of multiple dimensions, where $\Phi$ is not only a function of $r$ but also of other spatial coordinates, e.g. $\left(\theta,z\right)$ in the cylindrical coordinate system, $\left(\theta,\phi\right)$ in the spherical one, the equivalent to eq.~\ref{eq:mapping_deq} would be a set of differential equations that specify the mapping of the physical domain coordinate system to a reference one~\cite{brackbill1993an, fraggedakis2017discretization}. Integrating eq.~\ref{eq:mapping_deq} twice it follows that
\begin{equation}
    z' = \frac{C}{x^{(d-1)}}\rightarrow 
    z = \begin{cases}
 C\ln x \quad \text{for d=2} \\
 \frac{C}{x} \quad \text{for d=3}
 \end{cases}.
\end{equation}
Notice that these transformations between coordinates satisfy $$\nabla^2 z=0$$ in the respective dimensions: they are the natural coordinate in the curved geometry.
Without loss of generality we take $C=1$ and Eqn.\ref{eqn:OCP-radial-ode} now reads
\begin{equation}
 \begin{split}
  \partial_z^2 \Phi + e^{-\Phi+2z} & = 0 \quad \text{cylindrical} \\
  \partial_z^2 \Phi + \frac{ e^{-\Phi}}{z^4} & = 0 \quad \text{spherical} \\
 \end{split}
\end{equation}
No analytical solutions to the spherical case is known so far, but the cylindrical case can be recast into a similar form as in d=1, 
via the substitution $\psi=\Phi-2z$
\begin{equation}
 \begin{split}
  \partial_z^2 \psi + e^{-\psi} & = 0 \\
  \left( \frac{1}{2}\partial_z\psi \right)^2 - e^{-\psi} & = \xi
 \end{split}
\end{equation}
where the boundary condition is
\begin{equation}
 \begin{split}
  \partial_z \psi \left(z=\ln(r)\right) & = -2 \\
  \partial_z \psi \left(z=\ln(R)\right) & = R - 2
 \end{split}
\end{equation}
Assuming $\xi<0$ one arrives at
\begin{equation}
 \begin{split}
  \phi & = \frac{1}{Ze\beta} \ln \left[
  \frac{x^2}{\vert\xi\vert} \sin^2 \left( \sqrt{\frac{\vert\xi\vert}{2}} \ln(x) + \theta_0 \right) \right] \\
  & - \frac{1}{Z e \beta} \ln\left( \frac{q^2 e^2}{4\pi \epsilon k_B T c_0} \right)
 \end{split}
\end{equation}
with
\begin{equation}
 \begin{split}
  \sqrt{2\vert\xi\vert} \cot (\sqrt{\frac{\vert\xi\vert}{2}}\ln(r) + \theta_0) & = -2 \\
  \sqrt{2\vert\xi\vert} \cot (\sqrt{\frac{\vert\xi\vert}{2}}\ln(R) + \theta_0) & = R - 2 \\
 \end{split}
\end{equation}

\begin{figure*}
 \includegraphics[width=1\linewidth]{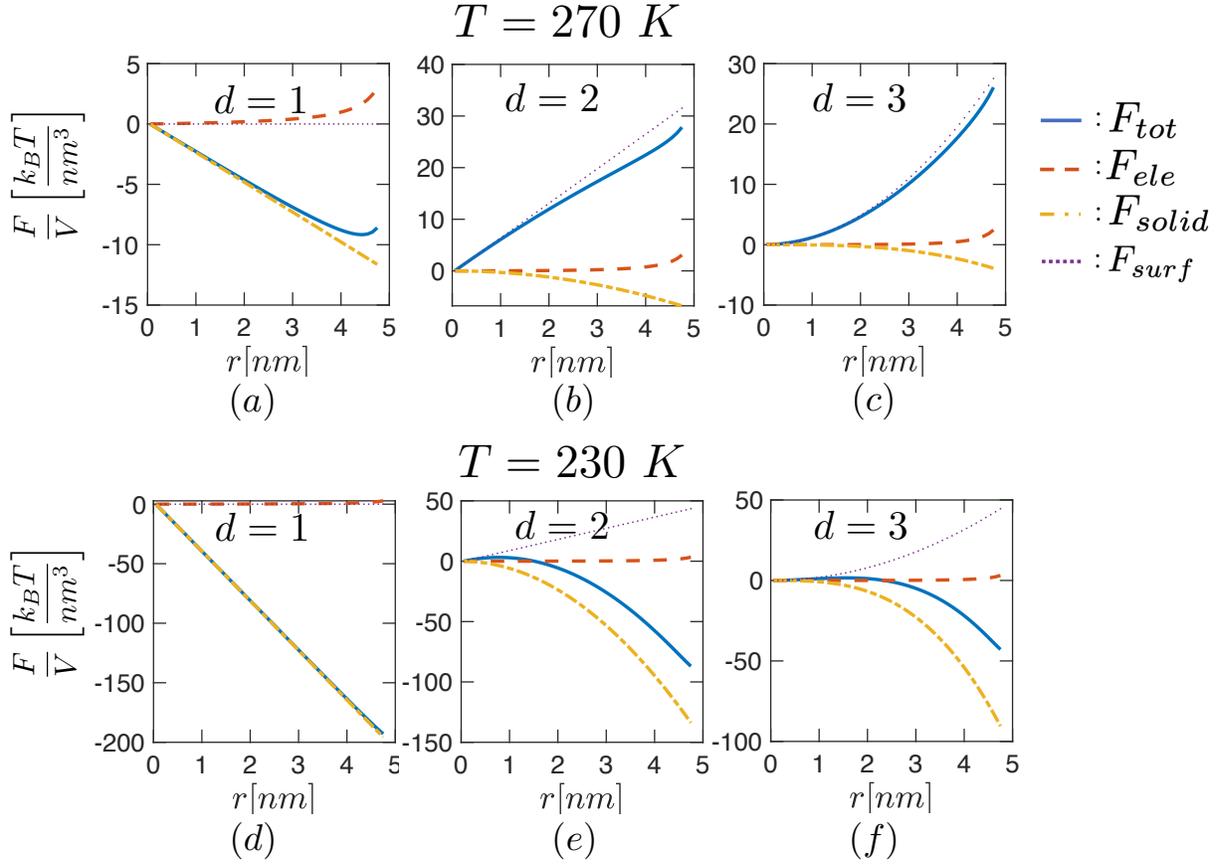}
 \caption{
  \label{fig:OCP-free-energy-profiles}
  The free energy for the case of freezing with free ions, as a function of $r$ in a pore of radius $R=5$~nm, with surface charge density $q=1$~nm$^{-2}$ and bulk water permitivity. Legend subscripts denotes the total free energy ($F_{tot}$), the electrolyte contribution ($F_{ele}$), the ice freezing enthalpy contribution ($F_{solid}$) and the surface tension term ($F_{surf}$). (a)(b)(c) at temperature $T=270$~K, (d)(e)(f) at $T=230$~K. (a)(d) d=1, (b)(e) d=2, (c)(f) d=3. Counter-ion recombination with pore surface charge is neglected. The equilibrium solution is given by the global minimum of $F_{tot}$ at $r^*$. If $r^*=0$ no freezing happens, otherwise $r^*>0$ freezing starts in the pore. For d=1, the pore almost freezes completely even at $T=272$~K, showing very minimal freezing point depression. For d=2,3 the competition between surface tension and freezing enthalpy overshadows the contribution from OCP. (a)(d)(e)(f) show frozen pores, while (b)(c) showing unfrozen pores. }
\end{figure*}

Surface tension in d=2,3 is not negligible: the typical values for pure water surface tension corresponds to $\sim 100$~MPa for 5~nm pore. By the above order of magnitude estimate on the pressure from OCP, we conclude that OCP contribution to $\Delta T$ and pressure is overshadowed by surface tension. The more interesting case of trapped salt ions in d=2,3 will be discussed later.
In Fig.\ref{fig:OCP-free-energy-profiles} (b)(e) and (c)(f) we show the free energy profiles for d=2,3. With the same parameters as in d=1, the effect of OCP is clearly overshadowed by the competition between surface tension and bulk freezing enthalpy.

\subsection{Trapped Ions Limit: 1:1 Electrolyte}
\label{sec:trapped ions}
In the case of nano-fluidic salt trapping, we focus on dealing with the electrolyte for the rest of this section.
For 1:1 electrolyte the equation reads
\begin{equation}
 \nabla^2 \Phi - \kappa^2 \sinh \Phi = 0
\end{equation}
where $\kappa^2=2 c_{ref} e^2 \beta/\epsilon$ and $c_{ref}$ is the concentration of salt at reference point of potential $\phi=0$, which is variable during freezing and compression of the electrolyte. In the limit of all salt ions being trapped the above Poisson-Boltzmann equation is constrained by number conservation of ions:
\begin{equation}
 c_{ref}(r) \int_r^R e^{-\Phi} S(d) x^{d-1} dx = N_i = \int_0^R c_0 e^{-\Phi_0} S(d) x^{d-1} dx
\end{equation}
The RHS is the total number of counter-ions when no ice has formed, and the LHS corresponds to the case where an ice core of radius $r$ has formed.

\subsubsection{d=1: 1:1 electrolyte in a slit pore geometry}
For 1:1 electrolyte the dimensionless PB equation reads
\begin{equation}
 \nabla^2\Phi - \sinh{\Phi} = 0
\end{equation}
by redefining the dimensionless variable $\tilde{x}=x/\lambda_D$, where the inverse Debye length is now $\kappa=\lambda_D^{-1}=\sqrt{2c_0 Z^2 e^2/\epsilon k_B T}$, and $\Phi=Z e \beta\phi$ the dimensionless potential. Again, for the ease of notation we remove the hat from all dimensionless quantities. The electrolyte concentration is determined by both the reference concentration $c_0$ and the electro-static potential $\phi$ as $c_i=c_0 e^{-\beta e Z_i \phi}$, with $Z_+ = -Z_- = Z$.
The boundary conditions for the 1:1 electrolyte are
\begin{equation}
 \begin{split}
  \Phi'(r) & = 0 \\
  \Phi'(R) & = -\lambda_D \frac{Ze^2q}{\epsilon k_B T} = -\frac{q}{2c_0 Z \lambda_D}
 \end{split}
 \label{eqn:salt-bc}
\end{equation}
where again $r$ corresponds to the location of the ice surface and $R$ is the radius of the charged pore. Additionally, charge neutrality leads to
\begin{equation}
 \begin{split}
  & \lambda_D^{d}\int_{r}^{R} S(d) x^{d-1}dx 2 Z c_0 \sinh{\Phi(x)} = - S(d) R^{d-1} \lambda_D^{d-1} q \\
  & \int_{r}^{R} x^{d-1}dx \sinh{\Phi(x)} =  - R^{d-1} \frac{q}{2 Z c_0 \lambda_D}
 \end{split}
\end{equation}

The nonlinear problem for $\Phi$ in general needs to be solved numerically. However, for the special case of $d=1$, an analytical solution can be derived that can reduce significantly the computational cost of evaluating the total free energy functional. 
Multiplying each side by $\Phi'$ we arrive at the following conserved form
\begin{equation}
 \frac{1}{2}\Phi'^2 - \cosh(\Phi) = \xi
 \label{eqn:salt-conservation-form}
\end{equation}
Then make a M\"{o}bius (fractional linear) transformation
\begin{equation}
 \begin{split}
  & \cosh(\Phi) = u = \frac{t+1}{at+1/a} \\
  & a^2+2\xi a + 1 = 0 \\
 \end{split}
\end{equation}
Eqn.\ref{eqn:salt-conservation-form} is rewritten as
\begin{equation}
 \frac{du}{\sqrt{(\xi+u)(u^2-1)}} = \pm \sqrt{2} dx
\end{equation}
where an additional change of variables leads to
\begin{equation}
 t = \frac{1-u/1}{au-1} = \frac{a-\cosh(\Phi)}{a^2\cosh(\Phi)-1}
\end{equation}
we arrive at
\begin{widetext}
 \begin{equation}
  \begin{split}
   \frac{du}{\sqrt{(\xi+u)(u^2-1)}} & = \frac{dt}{\sqrt{
    \left[ \left(1-1/a^2\right) + \left(1-a^2\right) t^2 \right]
    \left[ \left(1/a^2\xi-1/a\right) + \left(a^2\xi - a\right) t^2 \right]
    }} \\
   & = \frac{a^2 dt}{\sqrt{\left(a^2-1\right)\left(\xi-a\right)}
    \sqrt{
     \left[ 1 + \frac{a^2-a^4}{a^2-1} t^2 \right]
     \left[ 1 + \frac{\xi a^4-a^3}{\xi-a} t^2 \right]
     }} = \pm \sqrt{2} dx \\
  \end{split}
 \end{equation}
 The solution to the integral
 \begin{equation}
  \int_{0}^{t=\frac{a-\cosh(\Phi)}{a^2\cosh(\Phi)-1}} \frac{dt}{
   \sqrt{
    \left[ 1 - a^2 t^2 \right]
    \left[ 1 + \frac{\xi a^4-a^3}{\xi-a} t^2 \right]
    }} = \pm \sqrt{2}\frac{\sqrt{\left(a^2-1\right)\left(\xi-a\right)} }{a^2} (x - x_0)
 \end{equation}
 can now be expressed by the first kind of incomplete elliptic integral $F(u,v)$
 \begin{equation}
  (x - x_0) = \frac{i}{\sqrt{2\left(a^2-1\right)\left(\xi-a\right)}}
  F\left(
  \arctan \left( ia\frac{\cosh(\Phi)-a}{a^2\cosh(\Phi)-1} \right), \sqrt{1+\frac{\xi a^2-a}{\xi-a}}
  \right)
 \end{equation}
\end{widetext}
The implicit formula for $\Phi$ can be used to compute $F_{total}$ at each value of $r$. 

\subsubsection{d=2,3: Debye-H\"{u}ckel (DH) approximation in a cylindrical/spherical pore geometry}
In order to derive analytical solutions for 1:1 electrolytes in the case of d=2 or 3, we make the well-established Debye-H\"{u}ckel (DH) approximation~\citep{debye1923theorie}, which linearizes the exponential term of the PB equation. The DH equation for 1:1 electrolyte reads in dimensionless form
\begin{equation}
 \left( \nabla^2 - 1 \right) \Phi = 0
 \label{eqn:nondim-DH}
\end{equation}
where all lengths are again scaled by the Debye length $\lambda_D$.

Under cylindrical symmetry d=2, the radial part of Eqn.\ref{eqn:nondim-DH} is a Bessel equation
\begin{equation}
 \Phi'' + \frac{1}{x} \Phi' - \Phi = 0
 \label{eqn:DH-2D}
\end{equation}
Since angular dependency vanishes due to symmetry, the solution is constructed by
\begin{equation}
 \Phi = A j_0 (x) + B n_0 (x)
 \label{eqn:DH-2D-sol-general}
\end{equation}
where $j_0(x)$ and $n_0 (x)$ are the 0th order Bessel and Neumann functions, respectively. The coefficients $A$ and $B$ are determined by the given boundary condition of the problem.

Under spherical symmetry d=3, the radial part of Eqn.\ref{eqn:nondim-DH} gives a spherical Bessel equation
\begin{equation}
 \Phi'' + \frac{2}{x} \Phi' - \Phi = 0
 \label{eqn:DH-3D}
\end{equation}
and again, due to angular symmetry, we have
\begin{equation}
 \Phi = A \frac{e^{-x}}{x} + B \frac{e^x}{x}
 \label{eqn:DH-3D-sol-general}
\end{equation}
The coefficients $A$ and $B$ are given by the boundary conditions.
To remind us, the total free energy reads
\begin{equation}
 \begin{split}
  F & = (\mu_s - \mu_l) V(d) r^d + S(d) \gamma_{sl} r^{d-1} + \\
  & \int_{r}^{R} S(d) x^{d-1} dx
  \left( g(c) + \frac{\epsilon}{2} \phi'^2
  \right)
 \end{split}
\end{equation}
Minimizing $F$ w.r.t $r$ and noticing $\Phi'(x=r) = 0$ leads to a transcendental equation for $r$.

\begin{figure*}
 \includegraphics[width=1\linewidth]{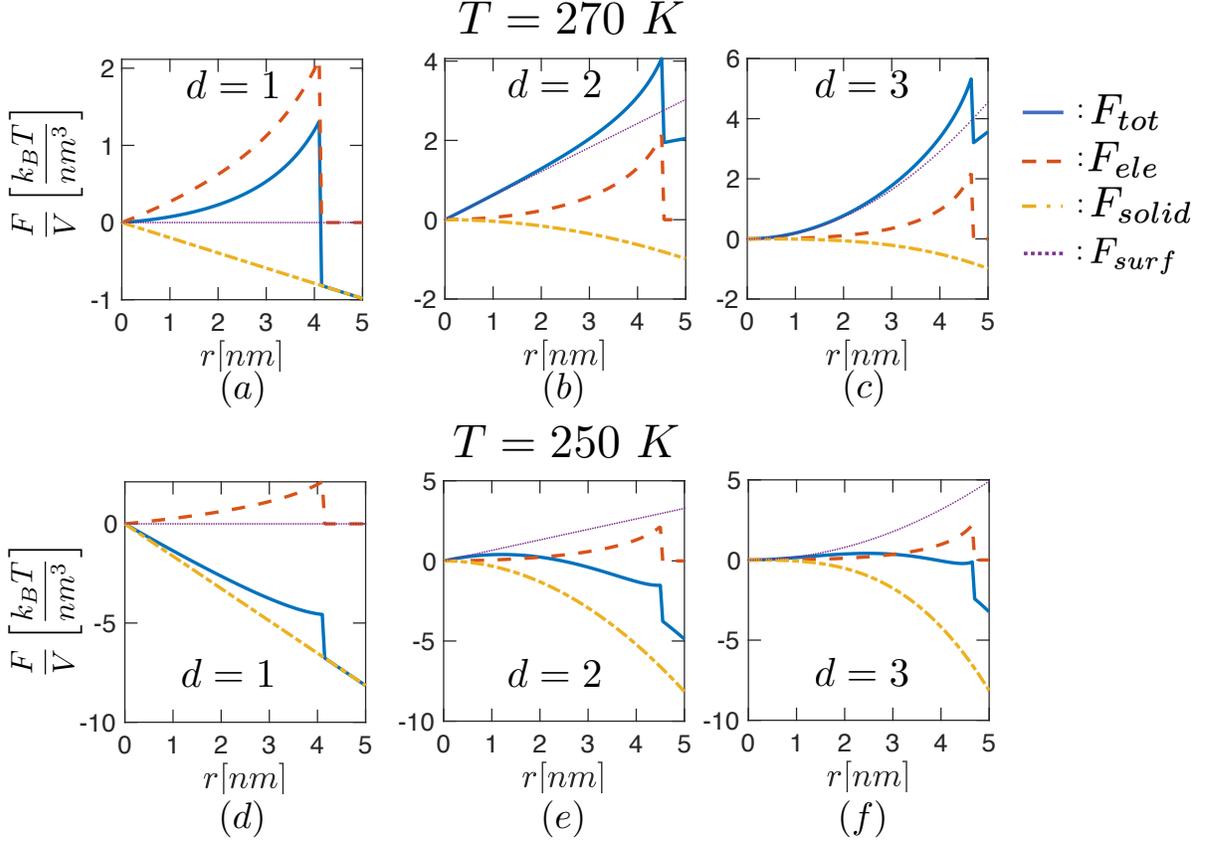}
 \caption{
  \label{fig:electrolyte-free-energy-profiles}
  The free energy for the case of freezing with nano-fluidic trapped ions, as a function of $r$ in a pore of radius $R=5$~nm, with surface charge density $q=1$~nm$^{-2}$ and assuming saturated water permitivity $\epsilon=10$. Initial salt concentrations for all subplots are $c_0=1$~M. Legend subscripts denotes the total free energy ($F_{tot}$), the electrolyte contribution ($F_{ele}$), the ice freezing enthalpy contribution ($F_{solid}$) and the surface tension term ($F_{surf}$). (a)(b)(c) at temperature $T=270$~K, (d)(e)(f) at $T=250$~K. Counter-ion recombination with pore surface charge is neglected. The equilibrium solution is given by the global minimum of $F_{tot}$ at $r^*$. If $r^*=0$ no freezing happens, otherwise $r^*>0$ freezing starts in the pore. The discontinuity in $F_{tot}$ and $F_{ele}$ happens at salt saturation. Beyond the saturation point $F_{tot}$ has no contribution from $F_{ele}$.
 (d)(e)(f) show frozen pores, while (b)(c) show unfrozen pores. Notice that in (a)  even though salt crystalization lead to a global minimum at the frozen state, the homogeneous nucleation barrier is sufficiently large that the unfrozen state is likely to be observed.}
\end{figure*}

\begin{figure}
 \includegraphics[width=1\linewidth]{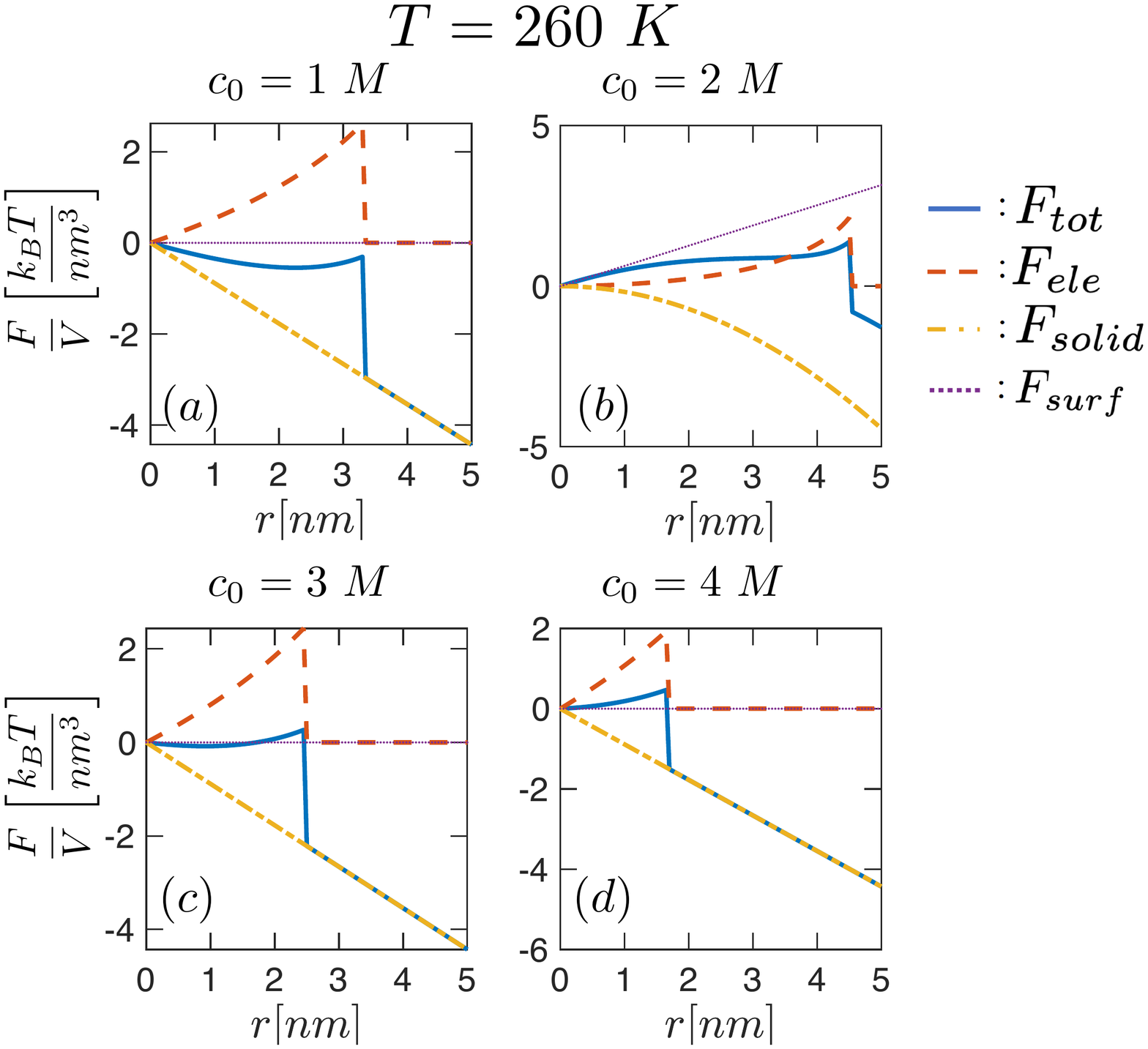}
 \caption{
  \label{fig:electrolyte-free-energy-profiles-concentration-dependence}
  The salt concentration dependence of free energy. Similar to Fig.\ref{fig:electrolyte-free-energy-profiles}, here we show the case of freezing with nano-fluidic trapped ions, as a function of $r$ in a pore of radius $R=5$~nm, with surface charge density $q=1$~nm$^{-2}$ and assuming saturated water permitivity $\epsilon=10$. Initial salt concentrations are (a) $c_0=1$~M, (b) $c_0=2$~M, (c) $c_0=3$~M, (d) $c_0=4$~M. Legend subscripts denotes the total free energy ($F_{tot}$), the electrolyte contribution ($F_{ele}$), the ice freezing enthalpy contribution ($F_{solid}$) and the surface tension term ($F_{surf}$). All subplots are at $T=260$~K for d=1. Counter-ion recombination with pore surface charge is neglected. The equilibrium state is given by the minimum of $F_{tot}$ before salt crystalization at $r^*$. If $r^*=0$ no freezing happens, otherwise $r^*>0$ freezing starts in the pore. Due to salt crystalization the global minima of (a-d) are all frozen states, however, the homogeneous nucleation barrier can be significant for the meta-stable states to be observed instead, in which case only (a) here is labled as fully frozen.}
\end{figure}

\subsection{Simplified Model for Thin Double Layer and Low Surface Charge}
While the mathematical expressions for $d=1,2,3$ are readily calculable numerically, greater physical intuition can be gained by considering the solution in the limit of small surface charge of the pore. In this limit, the entropic contribution to the free energy of the electrolyte becomes the $P-V$ work of compression of an ideal gas of ions with uniform concentration, valid when $\vert\Phi\vert\ll 1$. At small surface charge, the entropic contribution to changes in free energy dominate over the electrostatic contribution. The entropic contribution to the free energy change for $\vert\Phi \vert \approx 0$ is given by:
\begin{equation}
    F_{ent}=\int_{r}^{R} S(d) x^{d-1} dx
  \left( g(c_i) 
  \right)
\end{equation}
with uniform concentration for each species given by:
\begin{equation}
    c_+\approx c_-\approx  \frac{c_0  R^d}{(R^d-r^d)}
\end{equation}
The expression  for $F_{ent}$ can be easily integrated since $g(c_i)$ is a constant over the whole liquid domain.
\begin{equation}
    F_{ent}=\frac{2 k_B Tc_0 S(d) R^d}{d}\left(\ln(v c_0)-\ln\left(1-(r/R)^d\right)-1\right)
\end{equation}
As the ice core radius grows, the difference in free energy due to entropic effects versus $r=0$ is given by:
\begin{equation}\label{eq:eqEntropic}
    \Delta F_{ent}= -\frac{2 k_B T c_0 S(d)R^d }{d}\ln\left(1-(r/R)^d\right)
\end{equation}
The expression in Eq. \ref{eq:eqEntropic} can be expanded for $r/R\ll 1$ to give an additional contribution to the bulk energy of nucleation, which scales as $r^d$ at leading order.
\begin{equation}\label{eq:eqEntropicApprox}
    \Delta F_{ent}\approx \frac{2 k_B T c_0 S(d)r^d }{d}
\end{equation}
The dominant leading order approximation suggests that the pore size $R$ does not strongly affect the freezing point depression, unless the ice nuclei size $r$ is on the order of $R$. In other words, the freezing point depression will happen similarly in larger pores as long as the salt is trapped within the pore.

If the surface charge is fixed and the Debye length remains thin relative to the pore radius, then the electrostatic free energy will depend only on the change in the average concentration, captured by a change in the effective Debye length. The free energy stored in the diffuse part of the double layer at linear response is equal to the energy stored in a planar capacitor with capacitance $C=\kappa\epsilon$ multiplied by the outer pore surface area. The energy stored in the capacitor, in terms of the surface charge density, is:
\begin{equation}
    F_{field}=\frac{q^2 e^2 S(d) R^{d-1}}{2\epsilon\kappa}
\end{equation}
As the ice grows, the electrolyte is compressed, and the concentration of ions increases.  Therefore, the effective Debye length becomes smaller:
\begin{equation}
    \kappa(r)^2=\frac{1}{{\lambda_D(r)}^2}\approx\frac{2 c_0 e^2 \beta R^d}{\epsilon(R^d-r^d) }
\end{equation}
The capacitive energy decreases:
\begin{equation}
\Delta F_{field}= -\frac{q^2 e^2 S(d) R^{d-1}}{2\epsilon}\left(\frac{1}{\kappa(0)}-\frac{1}{\kappa(r)}\right) 
\end{equation}
or in terms of the ice core radius, $r$:
\begin{equation}
    \Delta F_{field}=-\frac{q^2 e^2S(d) R^{d-1}}{2\epsilon\kappa(0)}\left(1-\sqrt{1-(r/R)^d}\right)
\end{equation}
This expression can be expanded for $r/R\ll 1$ to give an additional contribution to the bulk free energy of freezing:
\begin{equation}\label{eq:eqFieldApprox}
    \Delta F_{field}\approx-\frac{q^2 e^2S(d) r^d}{4R\epsilon\kappa(0)}
\end{equation}
We can combine Eqs. \ref{eq:eqEntropicApprox} and \ref{eq:eqFieldApprox} to arrive at the modified bulk energy change due to ions:
\begin{equation}\label{eq:eqElecApprox}
\begin{split}
    \Delta F_{ele}&=\Delta F_{ent}+\Delta F_{field}\\&\approx \left(\frac{2 k_B T c_0 S(d) }{d}-\frac{q^2 e^2S(d)}{4R\epsilon\kappa(0)}\right)r^d
    \end{split}
\end{equation}

Note that the expression in Eq. \ref{eq:eqFieldApprox} scales with the square of the surface charge density, meaning at the low surface charge densities where the approximation is valid, the field energy change is negligible compared to the entropic change. While the simple formula gives a useful estimate of the free energy, geometrical confinement when the double layer is not thin renders the full numerical model necessary.  

These formulas can be used to gain intuition about the scales of the electrostatic and entropic contributions to the free energy profile as a function of $r$. They give simple modifications to the classical nucleation theory, as will be discussed in later sections. 

\subsection{Charge Regulation and Salt Crystallization}

When concentration of salt increases in the electrolyte, as ice forms, the surface charge will tend to recombine with counter-ions. The simplest recombination-dissociation equilibrium for a 1 step reaction with ion valence Z=1:
$\ce{M^{+} + B^- \rightleftharpoons MB}$ is described by
the equilibrium constant
$\ce{
 K=\frac{[M^{+}][B^-]}{[MB]}
}$.
The Langmuir adsorption isotherm for $M^+$ in OCP leads to
$s^2 q = \frac{\ce{K}}{\ce{K} + c_0 e^{-\Phi}}$,
where $s^2$ is the surface area occupied by a single site of $\ce{B^-}$ or $\ce{MB}$ group.  The boundary condition at the charged surface is then modified as:
\begin{equation}
 \Phi'(R) = \frac{e}{\epsilon s^2} \frac{K}{K + c_0 e^{-\Phi(R)}}
\end{equation}
The specific value of recombination-dissociation equilibrium constant depends on the chemistry of the pore surface charge and the counter-ion, and can be further complicated by the equilibrium of salt dissolution, multiple salt species and solution pH. Here above we present the general theory framework for it.

The analysis in Section.\ref{sec:trapped ions} is sufficient only if the salt is infinitely soluble in water. If the volume of electrolyte is reduced too much by ice formation such that the concentration of salt reaches saturation, the salt ions should precipitate into crystal. Once salt precipitation is triggered, the system becomes thermodynamically unstable and the new equilibrium will be complete freezing of the pore. In this state, all salt ions are transformed into crystal. This phenomenon is reflected in the discontinuities of the curves in Fig.\ref{fig:electrolyte-free-energy-profiles} and Fig.\ref{fig:electrolyte-free-energy-profiles-concentration-dependence}. We neglect the volume of salt crystal.

\subsection{Homogeneous Nucleation Barrier}
\begin{figure}
\centering
\begin{subfigure}{1\linewidth}
 \includegraphics[width=1\linewidth]{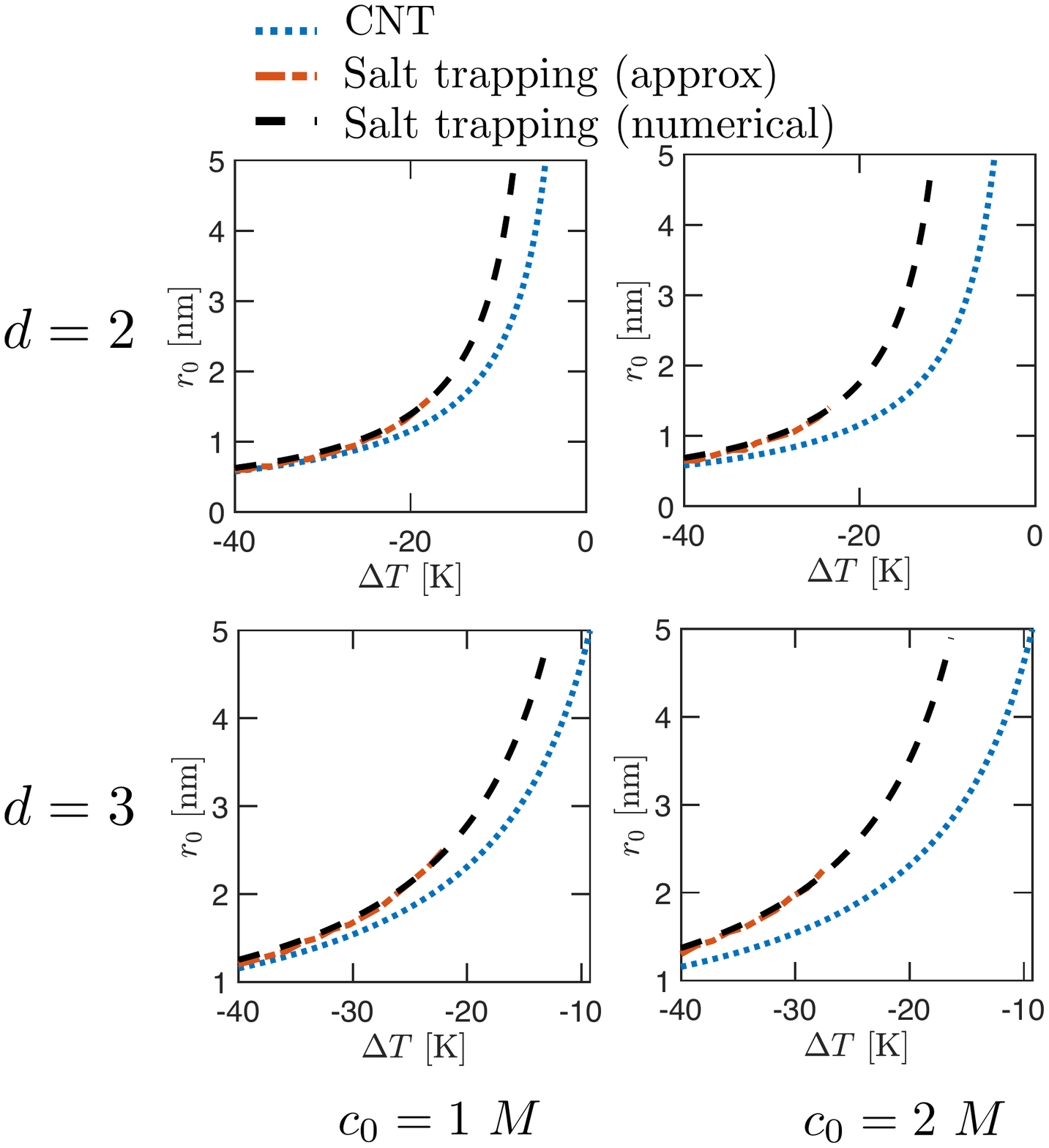}
\end{subfigure}
 \begin{subfigure}{1\linewidth}
 \includegraphics[width=1\linewidth]{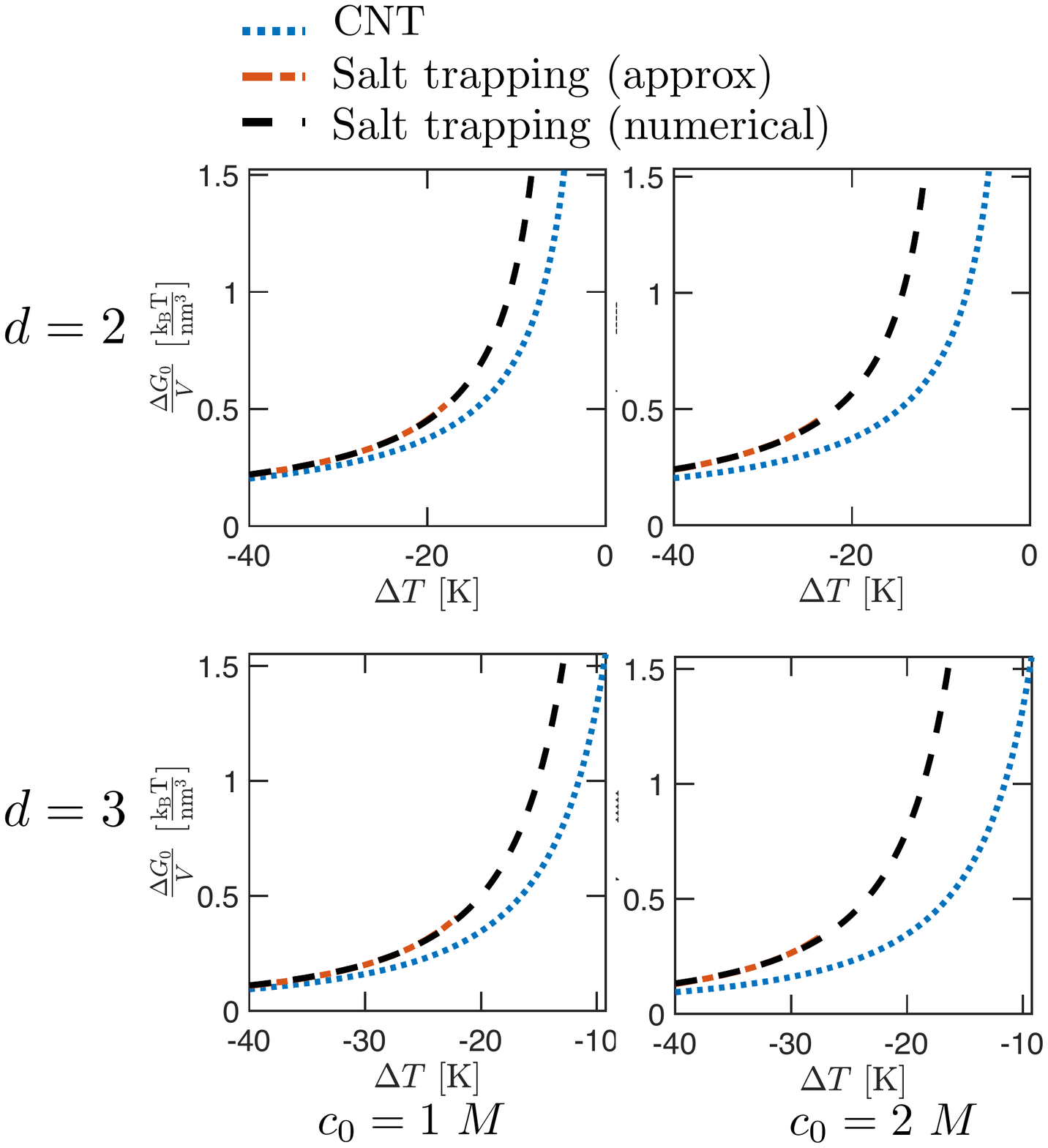}
\end{subfigure}
 \caption{
  \label{fig:nucleation}
  The critical nuclei radius ($r_0$) and energy barrier for nucleation ($\Delta G_0$) from classical nucleation theory, the simplified model (using Eq. \ref{eq:eqMuEff}) , and numerical solutions (based on Eq. \ref{eq:eqMinim}) for $d=2$ and $d=3$. The concentration is 1 mol/L salt and the surface charge density is 0.1 nm$^{-2}$, and the pore radius is $R=5$ nm. The critical nuclei radius and the energy barrier are both increased in the presence of trapped salt. The numerical solution is truncated when trapped salt makes freezing thermodynamically unfavorable (neglecting saturation of salt).
  }
\end{figure}

Now we consider effects of geometrical curvature, which imposes a homogeneous nucleation barrier due to solid-liquid interfacial tension, even in the absence of ions. In this section we still use the symbol $\Phi$ for dimensionless potential, and $R$ for dimensionless pore size, but restore $r$ to be of length dimension to denote critical nucleation radii. As shown in Fig \ref{fig:nucleation}, trapped salt again plays a crucial role by significantly increasing the free energy barrier and the critical radius of an ice nucleus forming in the pore.
Without ions, the critical nucleation radius $r_0$ and the nucleation free energy barrier for pure water confined in pores is described by the classical nucleation theory (CNT) as a result of competition between surface tension and bulk phase transformation:
\begin{equation}
\label{eqn:classical-nucleation}
 \begin{split}
  r_0 & = \frac{\gamma (d-1)}{\Delta\mu} \\
  \frac{\Delta G_0}{V} & = \gamma^d \left( \frac{d-1}{\Delta\mu} \right)^{d-1}
 \end{split}
\end{equation}
where $\Delta \mu \approx \vert Q \Delta T/T_0 \vert$ depends on the latent heat of freezing at bulk freezing point and the supercooling temperature.
In the presence of electrolyte the CNT equations are modified. 
Once we know the supercooling nucleation energy barrier as a function of temperature, given initial salt concentration and a certain energy threshold value we will know what the supercooling temperature is. 

As a first approximation, we can include the free energy change from the model assuming small surface charge density and thin double layers in the limit of $r/R\ll 1$, given by Eq. \ref{eq:eqElecApprox}. While this model does not capture the nonlinear complexity of the
problem, it can output a simple formula for the influence of salt trapping on ice nucleation. Because the terms to leading order are proportional to the ice core volume $\sim r^d$, they can be incorporated into an effective chemical potential change between the ice and water phases, $\Delta \mu_{eff}$:
\begin{equation}\label{eq:eqMuEff}
    \Delta \mu_{eff}\approx \vert Q \Delta T/T_0 \vert-2 k_B T c_0 +\frac{q^2 e^2 d}{4R\epsilon\kappa}.
\end{equation}
The value of $\Delta \mu_{eff}$ can be plugged into the expressions for $r_0$ and $\Delta G_0$ in place of $\Delta \mu$. The entropic contribution increases $r_0$ and $\Delta G_0$, whereas the field contribution acts to decrease them.

Now, we turn to calculating the nucleation barrier from the full numerical solution to the nonlinear model. Mathematically the nucleation barrier problem can be formulated as finding the extrema of the total free energy functional, $\Delta G_0$, that occurs at the critical ice core radius $r_0$. When performing this task, the complication introduced by the electrolyte reveals itself as below in the constrained optimization problem for the electrolyte region, which can be solved for the free energy density of electrolyte and is related to solving the PB equations with salt ion number conservation
\begin{equation}\label{eq:eqMinim}
    \begin{split}
            f(\phi(x;r,R),c(x;r,R)) & = \text{argmin} \int_r^R S(d)x^{d-1}dx f \\
            \text{s.t.} \nabla^2 \phi = Z e (c_+ - c_-) & = Ze c_0 \sinh(-\beta \phi)\\
            \phi'(r) & = 0\\
            \phi'(R) & = eq/\epsilon \\
            \int_r^R S(d) x^{d-1} c(x) dx & = N_{ion}
    \end{split}
\end{equation}
The free energy density $f$ of electrolyte modifies Eqn.\ref{eqn:classical-nucleation}.

We compare the results of the full numerical solution to the simple formula using $\Delta \mu_{eff}$ in Fig.\ref{fig:nucleation}. We find that the model works relatively well up to charge densities of the order $\sim$ 0.1~nm$^{-2}$.
Using transitioin state theory, the nucleation timescale can be calculated as
\begin{equation}
 \tau_{nuc} (\Delta T) = \tau_{0} e^{ \beta\Delta G_{ele}(\Delta T)}
\end{equation}
The lower panels of Fig.\ref{fig:nucleation} show the scaled nucleation timescale $\tau_{nuc}/\tau_0\propto\beta\Delta G_{ele}$ as a function of $\Delta T$. Since salt concentration increases the nucleation barrier in addition to surface tension effect, a longer nucleation time is expected.

\subsection{Gibbs-Thomson Effect of Melting Point Shift}
For d=2,3, the classical Gibbs-Thomson effect relates the shift of equilibrium melting point to the surface curvature, which is determined by the pore size in our context
\begin{equation}\label{eqn:gibbsthomson}
    \Delta T_m \propto \frac{1}{R}
\end{equation}
As shown in Fig.\ref{fig:electrolyte-free-energy-profiles}, after the salt saturates and crystallizes, the free energy does not admit any more contribution from the salt ions.  Hence, as the temperature goes up, the free energy of the frozen state and its variational neighborhood states, i.e when a little melting has initiated yet no salt ions can dissolve due to the pressure unfavorable of dissoluttion entropy, are always determined by the competition between surface tension and solidification enthalpy. Hence, the prediction of melting point will coincide with the classical Gibbs-Thomson effect (Eqn. \ref{eqn:gibbsthomson}).  

\section{Numerical Methods}
\subsection{One-Component Plasma}
We adopt an iterative method to solve Eqn.\ref{eqn:OCP-radial-ode} by expansion of $\Phi_t-\Phi_{t+1}$
\begin{equation}
 \Phi_{t+1}''+\frac{d-1}{x}\Phi_{t+1}'+e^{-\Phi_{t}}e^{\Phi_t-\Phi_{t+1}} = 0
\end{equation}
which after keeping the first order term of $e^{\Phi_t-\Phi_{t+1}}\approx 1 + \Phi_t-\Phi_{t+1}$ and manipulation of terms we get the conservative form
\begin{equation}
 \partial_x x^{d-1}\Phi_{t+1}'(x) - x^{d-1} e^{-\Phi_{t}}\Phi_{t+1} = -(1+\Phi_t) x^{d-1} e^{-\Phi_{t}}
 \label{eqn:numerics-OCP-iterative}
\end{equation}
Starting from a reasonable initial trial solution $\Phi_{t=0}\equiv 0$, Eqn.\ref{eqn:numerics-OCP-iterative} gives the rule for iteration, whose stopping criteria are
\begin{equation}
 \abs{\Phi_t-\Phi_{t+1}} < \delta_{iter} \qquad or \qquad N_{iter}>N_{itermax}
 \label{eqn:numerics-OCP-iterative-stopping-criteria}
\end{equation}
After $\Phi(x)$ is evaluated, the field energy can be integrated
\begin{equation}
 F_{field} = \mathcal{F} \int_{r}^{R} x^{d-1} dx \frac{1}{2}(\Phi')^2
 \label{eqn:field-energy-ocp}
\end{equation}
where $\mathcal{F}=k_B T \frac{S(d)}{4\pi} \frac{l_G^{d-2}}{l_B}$.
The ideal gas entropy is
\begin{equation}
 \begin{split}
  & -TS = k_B T l_G^d \int_{r}^{R} S(d) x^{d-1} dx c(x) \left[ \ln ( c(x) v ) -1  \right] \\
  & = \mathcal{F} \int_{r}^{R} x^{d-1} dx e^{-\Phi(x)} \left[ \ln\left( \frac{v}{4\pi l_B l_G^2} \right) -\Phi(x) - 1 \right]
 \end{split}
 \label{eqn:ideal-gas-entropy-ocp}
\end{equation}
where $v$ is counter-ion volume. In practice we take the volume of hydrated ions. Notice that $\Phi$ is usually positive from the numerical solution.
The free energy of ice is
\begin{equation}
 \begin{split}
  F_s & = (\mu_s-\mu_l) l_G^{d} V(d) r^d = \left( Q\frac{\Delta T}{T_0} + \Delta c_p \frac{\Delta T^2}{T_0} \right) l_G^d V(d)r^d \\
  & = \left( Q\frac{\Delta T}{T_0} + \Delta c_p \frac{\Delta T^2}{T_0} \right) \frac{4\pi l_G^2 l_B V(d)}{k_B T S(d)} r^d \mathcal{F}
 \end{split}
\end{equation}
where $V(1)=1, V(2)=\pi, V(3)=4\pi/3$.
The surface energy is
\begin{equation}
 \begin{split}
  F_{surf} & = S(d) r^{d-1} l_G^{d-1} \gamma_{sl} \\
 \end{split}
\end{equation}

The problem now is reduced to finding the minimum of the functional w.r.t. $r$
\begin{equation}
 \begin{split}
  & \int_{r}^{R} x^{d-1} dx \left\{ \frac{1}{2}(\Phi')^2 + v e^{-\Phi(x)} \left[ \ln\left( \frac{v}{4\pi l_B l_G^2} \right) -\Phi(x) - 1 \right] \right\} \\
  & + \left( Q\frac{\Delta T}{T_0} + \Delta c_p \frac{\Delta T^2}{T_0} \right) \frac{4\pi l_G Z V(d)}{q k_B T S(d)} r^d + 4\pi r^{d-1} \frac{Z \gamma_{sl}}{k_B T q}
 \end{split}
\end{equation}
where $\gamma_{sl}$ is the solid-liquid surface energy per unit area.


\subsection{Trapped Salts}
The same iterative algorithm as above is applied to trapped salts where the PB equation reads
\begin{equation}
\partial_x x^{d-1} \Phi'_{t+1} - x^{d-1} \Phi_{t+1} \cosh\Phi_t = x^{d-1} \left(
\sinh\Phi_t - \Phi_t \cosh\Phi_t
\right)
\end{equation}
with $x$ the spatial coordinate scaled by the Debye length $\lambda_D$. 

Now the field energy is similar to Eqn.\ref{eqn:field-energy-ocp}
\begin{equation}
 F_{field} = \mathcal{F} \int_{r}^{R} x^{d-1} dx \frac{1}{2}c_0 (\Phi')^2
 \label{eqn:field-energy-pb}
\end{equation}
except now $\mathcal{F}=2S(d)\lambda_D^d k_B T$, and $r,R$ are normalized by $\lambda_D$.
The ideal gas entropy is summing over 2 ion species
\begin{equation}
 \begin{split}
  & -TS = k_B T \lambda_D^d \int_{r}^{R} S(d) x^{d-1} dx \sum_{i=1,2} c_i(x) \left[ \ln ( c_i(x) v ) -1  \right] \\
  & = \mathcal{F} \int_{r}^{R} x^{d-1} dx c_0 \left[
  \left( \ln vc_0 - 1 \right) \cosh \Phi + \Phi \sinh \Phi
  \right]
 \end{split}
 \label{eqn:ideal-gas-entropy-pb}
\end{equation}
The free energy of ice is normalized by $\lambda_D$ as
\begin{equation}
 \begin{split}
  F_s & = (\mu_s-\mu_l) \lambda_D^{d} V(d) r^d = \left( Q\frac{\Delta T}{T_0} + \Delta c_p \frac{\Delta T^2}{T_0} \right) \lambda_D^d V(d)r^d \\
  & = \left( Q\frac{\Delta T}{T_0} + \Delta c_p \frac{\Delta T^2}{T_0} \right) \frac{ V(d)}{2 k_B T S(d)} r^d \mathcal{F}
 \end{split}
\end{equation}
The surface energy is
\begin{equation}
 F_{surf} = S(d) r^{d-1} \gamma_{sl} \lambda_D^{d-1} = \mathcal{F} \frac{\gamma_{sl}}{2\lambda_D k_B T} r^{d-1}
\end{equation}

Similar to Eqn.\ref{eqn:numerics-OCP-iterative}, we use the first order expansion of $\sinh(\Phi_t+\Delta)\sim \sinh(\Phi_t)+\Delta \cosh(\Phi_t)$ for an iterative scheme.
After solving this finite difference problem one can minimize the functional below
\begin{equation}
 \begin{split}
  & \int_{r}^{R} x^{d-1} dx c_0 \left( \frac{1}{2} \Phi'^2 + \cosh\Phi \left( \ln vc_0 - 1 \right) + \Phi \sinh\Phi \right) + \\
  & \left( Q\frac{\Delta T}{T_0} + \Delta c_p \frac{\Delta T^2}{T_0} \right) \frac{ V(d)}{2 k_B T S(d)} r^d + \frac{\gamma_{sl}}{2\lambda_D k_B T} r^{d-1}
 \end{split}
 \label{eqn:main-minimization}
\end{equation}

\section{Conclusions and Discussions}

In this paper, we present a general continuum theoretical framework to model freezing phenomena in charged heterogeneous porous media.  We distinguish the regimes of free and trapped salt ions, which can rise from bottlenecks and heterogeneous freezing, referred as nano-fluidic trapping. The limit of free ions is approximated as one-component plasma, and is shown to only induce minimal freezing point depression and pressure. While in the case of trapped salts, the freezing process becomes continuous, distinct from the feature of bulk freezing as a first order phase transition. The freezing point depression and pressure are significant in typical situations of biological or material science applications, which we elaborate in another companion paper. We discuss and include in the framework additional physical chemistry phenomena such as charge regulation and salt saturation. Finally, both numerical results and analytical approximations are obtained to derive a modified nucleation theory, when surface tension effects are combined with the influence of the trapped salt ions.
Our theory can find potential applications in freezing tolerance/endurance of biological and inorganic materials, or novel nano-fluidic devices.

For the sake of simplicity and demonstration, some approximations have been made throughout this paper, which could be investigated and potentially relaxed in future analysis. Ice formation is treated as a homogeneous nucleation process with a stable growing interface, where heterogeneous nucleation or dendritic growth regimes are possible. A stability analysis on the growth interface could clarify the regime of validity of this assumption. Solubility of salt ions in ice is neglected, which implies a not too fast freezing process, and could be violated in a very fast vitrification arising from large supercooling. Lastly, continuum predictions may deviate from reality when the freezing pores are reaching the size of nm; for example the discrete size of solvated ions may have to be considered. 

\acknowledgments{
T. Zhou thanks S. Yip, M. Pinson and Z. He for helpful discussions. T. Zhou is grateful for the support of Concrete Sustainability Hub (CSHub) at MIT. 
}

%

\end{document}